\documentclass[prd,twocolumn,nofootinbib,showpacs,amsmath,amsfonts]{revtex4}

\usepackage{graphicx}

\newcommand*{\fig}[1]{Fig.~\ref{#1}}

\newcommand*{\pol}{\langle |L| \rangle}
\newcommand*{\zpol}{\langle |L_P| \rangle}
\newcommand*{\Zt}{\mathrm{Z}_2}

\begin{document}

\title{Center Dominance in SU(2) Gauge-Higgs Theory}

\date{\today}

\author{Roman Bertle}
  \email{bertle@kph.tuwien.ac.at}
\author{Manfried Faber}
  \email{faber@kph.tuwien.ac.at}
\affiliation{%
Atomic Institute of the Austrian Universities,
Nuclear Physics Division,
A-1040 Vienna,
Austria}

\author{Jeff Greensite}
  \email{greensit@stars.sfsu.edu}
\affiliation{Physics and Astronomy Dept.,
San Francisco State University,
San Francisco,
CA 94117,
USA}

\author{{\v S}tefan Olejn\'{\i}k}
  \email{stefan.olejnik@savba.sk}
\affiliation{Institute of Physics,
Slovak Academy of Sciences,
SK-845 11 Bratislava,
Slovakia}

\begin{abstract}
We study the SU(2) gauge-Higgs system in $D=4$ dimensions, and analyze the influence of the fundamental-representation Higgs field on the vortex content of the gauge field.
It is shown that center projected Polyakov lines, at low temperature, are finite in the infinite volume limit, which means that the center vortex distribution is consistent with color screening.
In addition we confirm and further investigate the presence of a ``Kert\'esz line'' in the strong-coupling region of the phase diagram, which we relate to the percolation properties of center vortices.
 It is shown that this Kert\'esz line separates the gauge-Higgs phase diagram into two regions: a confinement-like region, in which center vortices percolate, and a Higgs region, in which they do not.
 The free energy of the gauge-Higgs system, however, is analytic across the Kert\'esz line.
\end{abstract}


\maketitle

\section{\label{sec:intro}Introduction}

Lattice studies strongly support the vortex picture of confinement and the importance of the center degrees of freedom (see review \cite{gr03a} and references therein).
Until recently, these numerical investigations concerned pure gluonic QCD only.
If dynamical fermions are included, it is well known that the long range part of the potential between static charges changes qualitatively:
the string breaks and the potential levels off at a screening distance $r_0$.
String breaking has been observed numerically not only for gauge fields with dynamical fermions, but also for gauge-Higgs theories \cite{ks98a}.
In this work we investigate the influence of dynamical matter fields on the distribution of center vortices.
Our model is lattice SU(2) gauge-Higgs theory with Higgs fields in the fundamental representation.

The SU(2) gauge-Higgs model is defined by the action

\begin{align}
  \label{eq:action}
  S &= S_W 
    \sum_x \left[\Phi^\dagger(x)\Phi(x) +
    \lambda\left(\Phi^\dagger(x)\Phi(x)-1\right)^2\right] \notag\\
    &- \kappa \sum_{\mu,x} \left(\Phi^\dagger(x)U_\mu(x)\Phi(x+\hat{\mu}) +
    \text{cc}\right)\\
S_W &= \beta \sum_{\mu<\nu,x} \left(1-\frac{1}{2}\text{Re}\,\text{Tr}\,
  U_{\mu\nu}(x)\right)\;,
\end{align}
where $S_W$ is the usual Wilson plaquette action, and $\Phi=\binom{\phi_1}{\phi_2}$, with $\phi_1, \phi_2 \in \mathbb{C}$, is a massive scalar field in the fundamental representation.

\begin{figure}
  \includegraphics[width=0.7\linewidth,height=0.3\textheight,keepaspectratio]{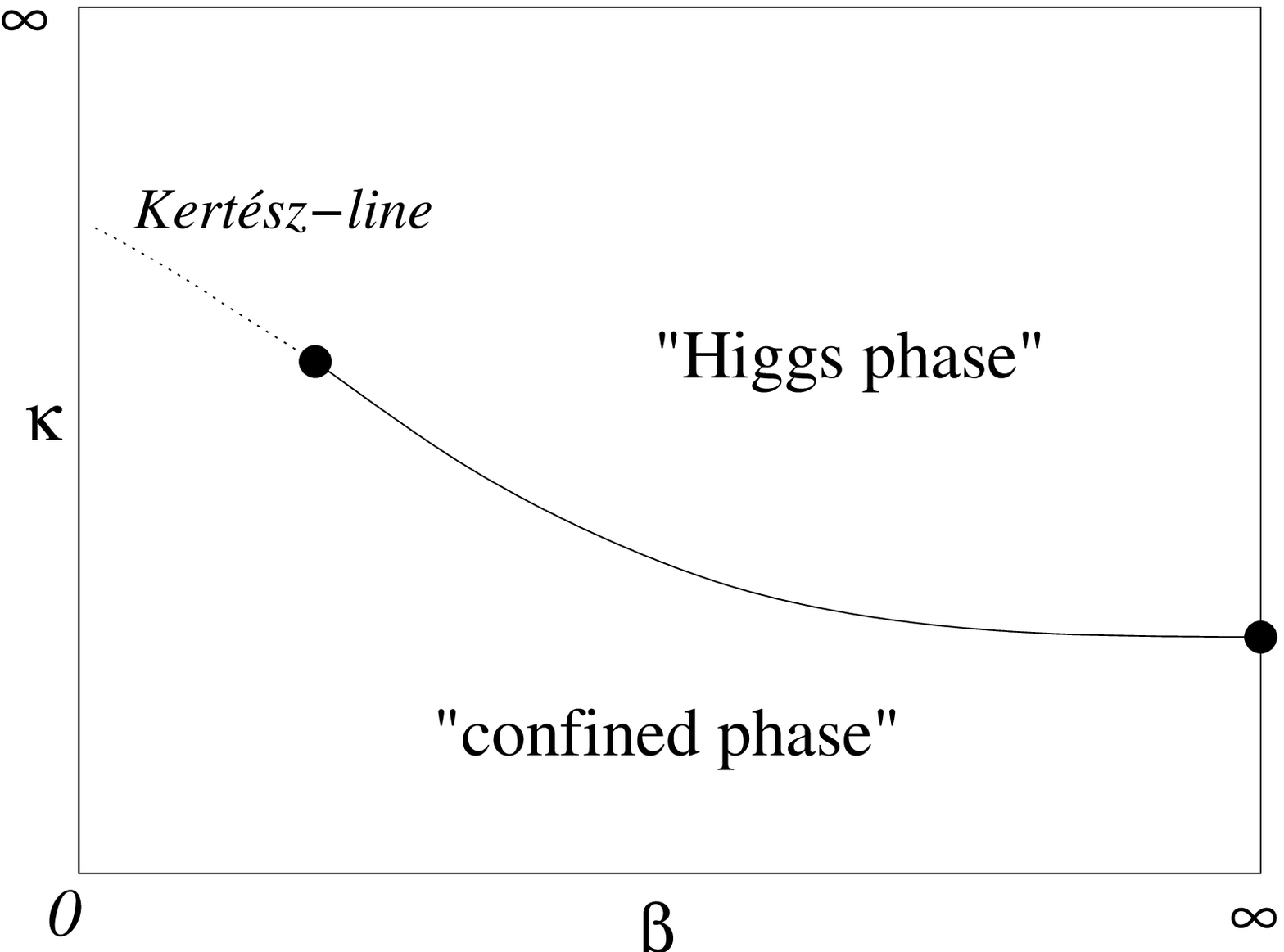}\\[2ex]
  \includegraphics[width=0.7\linewidth,height=0.3\textheight,keepaspectratio]{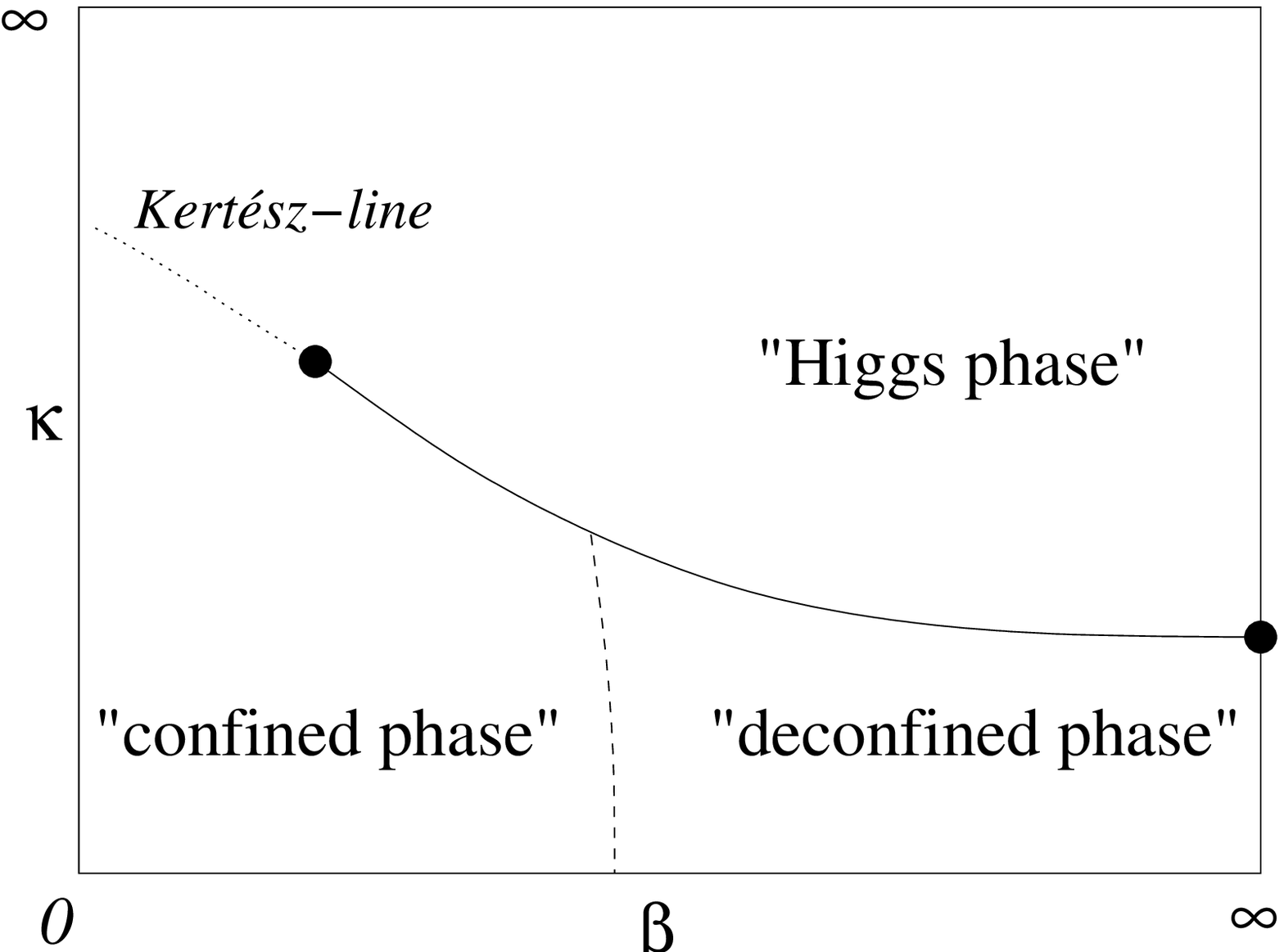}
  \caption{Schematic phase diagram for the SU(2)-Higgs system at zero temperature (top), and on lattices with a fixed extension in the time direction (bottom).}
  \label{fig:phasediag}
\end{figure}

A schematic phase diagram for the SU(2)-Higgs model is depicted in
\fig{fig:phasediag}.
In the ``confined'' phase, the potential between fundamental charges rises linearly at intermediate distances.
Due to color screening of fundamental charges, there is string breaking at some finite distance $r_0$, and the potential levels off.
In the Higgs phase, the Higgs mechanism is at work, and the potential is Yukawa-like;
the string tension vanishes at all separations.
However, these are not thermodynamically distinct phases.
The phase diagram is connected, in the sense that one can always find a path between any two points in the phase diagram which avoids any non-analyticity in thermodynamic quantities.
The transition line which might have separated the Higgs and confinement phases ends at a critical point in the strong coupling region, and from there changes over
to crossover behaviour.
Thus if we speak of the ``Higgs'' or the ``confinement'' phases in this model, we are aware that we speak rather of regions belonging to a single phase of the system \cite{fs79a}.

At finite temperature and $\kappa=0$ there is an additional phase, namely, the deconfined phase.
In this phase the quark-gluon plasma arises and fundamental charges are set free.
For $\kappa>0$ there is no true phase transition between the ``confined'' and ``deconfined'' phases, but only crossover behavior indicated by the dashed line in \fig{fig:phasediag} (lower figure).
Hence we still have only one phase in the $\beta-\kappa$ plane.

The vortex theory of confinement was put forward at the end of the 1970s (see references in review \cite{gr03a});
early applications to SU(2) gauge-Higgs theory  can be found in refs.\ \cite{mm82a,me84b}.
According to the vortex picture, tube-like (D=3) or surface-like (D=4) objects, carrying quantized amounts of magnetic flux, play a crucial role.
The standard procedure to identify these objects (the center vortices) on the lattice is to first gauge-fix the lattice configurations according to a gauge-fixing condition on link variables in the adjoint representation.
Then the center degrees of freedom are singled out by projecting each SU(2) group-valued link variable to the closest $\Zt$ center element.
In this work we use the direct maximal center gauge (DMCG) \cite{dfggo98a} fixed by the over-relaxation method.
Maximizing $\langle|\mathrm{Tr}[U_\mu(x)]|^2\rangle$, DMCG shifts the link variables $U_\mu(x)$ as close as possible towards the center elements of SU(2).
Center projection in DMCG has proven to be a useful tool for isolating the relevant degrees of freedom on the lattice, although other adjoint gauges, and more sophisticated gauge-fixing techniques, are also in use.
Via center projection, the SU(2) lattice is mapped to a $\Zt$ gauge field configuration.
The excitations of $\Zt$ lattice configurations are the P-vortices.
P-vortices form closed surfaces (in four dimensions) or closed loops (in three dimensions) on the dual lattice, and are composed of P-plaquettes.
A plaquette on the projected lattice is said to be pierced by a P-vortex, and is called a ``P-plaquette'', if the product of the four projected links of the plaquette yields the nontrivial element of $\Zt$, i.e.\ $-1$.
Center projection in an adjoint gauge can be viewed as a tool to locate the position of thick SU(2) vortices from the location of P-vortices on the projected lattice.

Here we are interested in the dependence of the center vortex distribution on couplings and temperature.
The vortex model at finite temperature, and the behaviour of P-vortices at the transition to the high temperature phase, have been studied in recent years in refs.\ \cite{dfgo97b,cpvz98a,lter99a,elrt99a,bfgo99a}.
In those articles it was shown that center dominance $-$ i.e.\ the correspondence of projected and unprojected observables sensitive to infrared physics $-$ holds also at finite temperatures.
It was found \cite{cpvz98a,lter99a,elrt99a,bfgo99a} that the vortex density $p$ --- the fraction of plaquettes on the projected lattice which are P-plaquettes --- drops substantially across the deconfinement transition and, more importantly, that vortices do not percolate in any space-slice, i.e.\ there is no percolation of vortex lines in a three-volume with either the $x$, $y$, or $z$ coordinate held fixed.
 The absence of vortex percolation in a space-slice implies that center projected Polyakov lines have non-zero expectation values in the deconfined phase \cite{lter99a,elrt99a}.

Recent findings \cite{la02a,bf02a} indicate that the vortex density also decreases sharply in the gauge-Higgs model, at the transition from the ``confined'' phase to the Higgs phase.
This sharp decrease in vortex density occurs even in the crossover region at small values of $\beta$, where the phase transition line has ended.
It was suggested by Langfeld \cite{la02a} that the cross-over line at small $\beta$, where there is a sudden drop in the vortex density, could be a ``Kert\'esz line'' of the sort found in the Ising model \cite{ke89a, sa02a}.
A Kert\'esz line is a line of percolation transitions which is not associated with a thermodynamic transition, e.g.\ from an ordered to a disordered state.
 Our new calculations provide further support for the existence of a Kert\'esz line in the gauge-Higgs system.
Although there is no thermodynamic transition from a ``confined'' to a Higgs phase, and indeed (in contrast to pure gauge theory) there is no true confined phase in this system, we nonetheless find a line of center vortex depercolation.

\section{\label{sec:fintemp}Finite Temperature}

It is most efficient to carry out Monte Carlo simulations of the gauge-Higgs system in unitary gauge, where $\Phi=\binom{\phi}{0}, \phi \in [0,\infty]$, and only one degree of freedom has to be simulated for the Higgs field.
 In the unitary gauge we have
\begin{align}
  \label{eq:unipart}
    Z &= \int \mathcal{D}[U]\mathcal{D}[\phi]
  \exp\left(-S_W-S_\mathrm{Hu}\right)\\
  \label{eq:uniaction}
  \begin{split}
    S_\mathrm{Hu} &= \sum_x
      \left[\phi^2(x) + \lambda\left(\phi^2(x)-1\right)^2\right]\\
    &- \kappa \sum_{\mu,x} \Bigl(\phi(x)\phi(x+\hat{\mu})\mathrm{Tr}[U_\mu(x)]\Bigr)
  \end{split}
\end{align}
with $\phi = \sqrt{\Phi^\dagger \Phi}$ and integration measure
\begin{align}
 \int\mathcal{D}[\phi]=\prod_x\int_0^\infty
\mathrm{d}\phi(x)\phi^3(x).
\end{align}
We used the Metropolis algorithm for the numerical simulation of the system on $D=4$ dimensional Euclidean lattices of dimensions $N_t*N_s^3$.
Data was taken on lattices separated by 50 Monte Carlo steps.
For the measurement of center projected observables, we fixed to DMCG using the overrelaxation method and then carried out the projection.

At finite temperature, confinement and deconfinement can be studied using the Polyakov loop observable
\begin{equation}
\langle |L| \rangle = \Big\langle \Big| \frac{1}{N_s^3}
\sum_{\mbox{\bf x}} L(\mbox{\bf x}) \Big| \Big\rangle = e^{-F/T}
\text{ ~~for~~ } N_s\rightarrow\infty
\end{equation}
where 
\begin{equation}
  L(\mbox{\bf x}):= \frac{1}{2}\mathrm{Tr}\prod_{x_4}U_4(\mbox{\bf x}, x_4)
\end{equation}
and $F$ is the free energy of a single static quark relative to vacuum at temperature $T$.
Without dynamical matter fields, we have
\begin{equation}
\lim_{N_s\rightarrow\infty} \pol = 0
\end{equation}
in the confined low temperature phase.
If matter is included, screening becomes possible.
In that case $\pol$ is non-zero, and $F$ is finite, in the large volume limit.

\begin{figure}
  \includegraphics[width=1.0\linewidth,height=0.4\textheight,keepaspectratio]{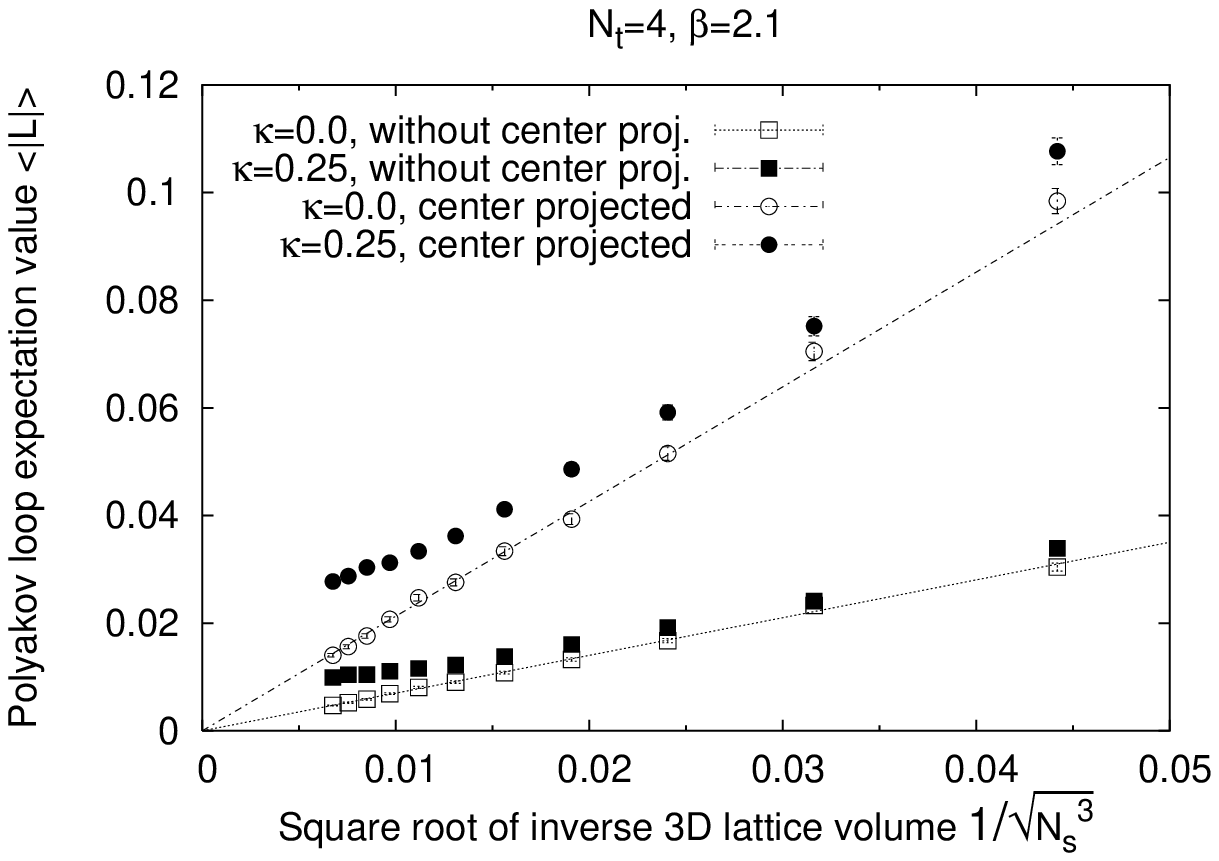}
  \includegraphics[width=1.0\linewidth,height=0.4\textheight,keepaspectratio]{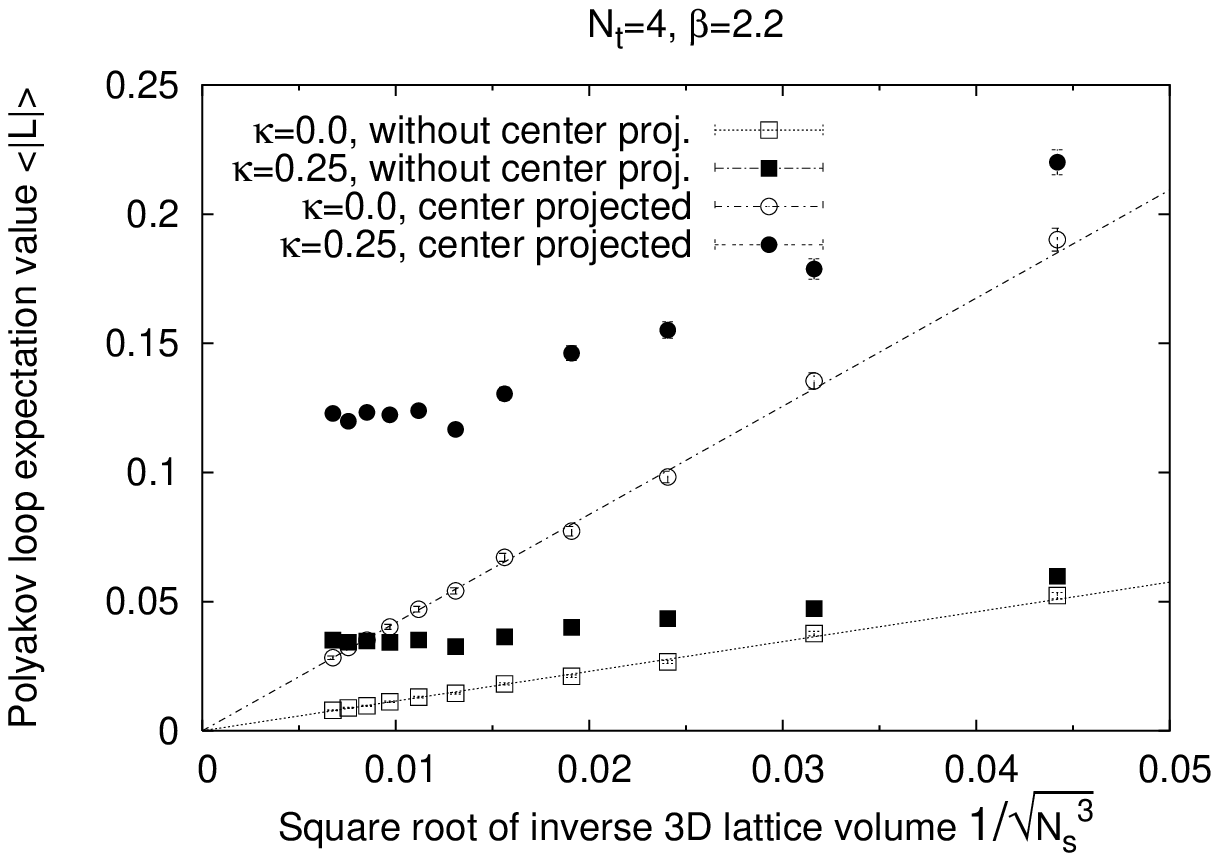}
  \caption{Expectation values of the Polyakov loop in the ``confined'' phase.
    Open symbols denote measurements at $\kappa=0.0$, filled symbols at $\kappa=0.25$.
    Squares are used for measurements on unprojected, full SU(2) configurations and circles for center projected fields.}
  \label{fig:polyakovconf}
\end{figure}

Since the idea of the vortex model is to describe the confining properties of gauge fields by center degrees of freedom, the question arises whether this observed screening can also be seen after singling out these variables.
We have carried out center projection in DMCG without ($\kappa=0$) and with ($\kappa=0.25$) scalar matter fields on lattices of size $N_t*N_s^3$ with $N_t=4$ and $N_s$ varying from 8 to 28.
The quartic coupling has been set to $\lambda=0.5$, and data has been taken on 1000 configurations for each parameter set.
Some results are shown in \fig{fig:polyakovconf} for $\beta=2.1$ and $2.2$, where the system is in the low temperature ``confined'' phase.

\begin{figure}
  \includegraphics[width=1.0\linewidth,height=0.4\textheight,keepaspectratio]{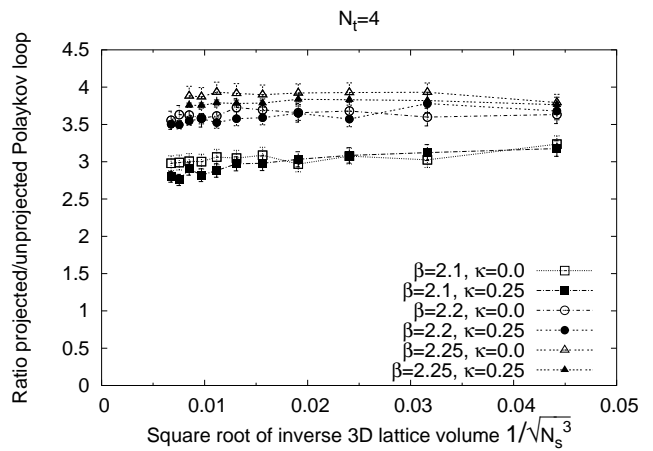}
  \caption{Ratio of projected to unprojected Polyakov loop expectation values in the ``confined'' phase.
    Open symbols denote measurements at $\kappa=0.0$, filled symbols at $\kappa=0.25$.}
  \label{fig:polyakovratioconf}
\end{figure}

At the chosen $\beta$ values and $\kappa=0$, the expectation value of $\pol$ vs.\ $1/\sqrt{N_s^3}$ on unprojected lattices extrapolates linearly to zero in the $N_s \rightarrow \infty$ limit of infinite spatial volume.
In contrast, at finite coupling $\kappa=0.25$, $\pol$ tends to a non-zero constant at infinite spatial volume.
This well known feature, associated with the screening of static charges by dynamical matter fields, is also found on center projected lattices.
Due to suppressed self energy contributions and ultraviolet fluctuations in the projected configurations, the expectation value of the Polyakov loop observable $\zpol$ on projected lattices is larger than the corresponding expectation value $\pol$ on unprojected lattices, but the qualitative behaviour of the two observables, in the large-volume limit, is the same.
In fact, the ratio of the projected and unprojected Polyakov expectation values is virtually constant, and nearly independent of both lattice extension $N_s$ and coupling $\kappa$ for all values of $\beta=2.1$, $2.2$ and $2.25$, as seen in \fig{fig:polyakovratioconf}.

At $\kappa=0$ the Polyakov expectation value $\pol$ is zero in the infinite volume limit at low temperature, and non-zero beyond the deconfinement transition.
This transition can be understood in the framework of the vortex model \cite{lter99a,elrt99a,cpvz98a,bfgo99a}.
The projected Polyakov line $\zpol$ has the same transition from zero to non-zero values at the deconfinement temperature, as already reported in ref.\  \cite{dfgo97b}.

\begin{figure}
  \includegraphics[width=1.0\linewidth,height=0.4\textheight,keepaspectratio]{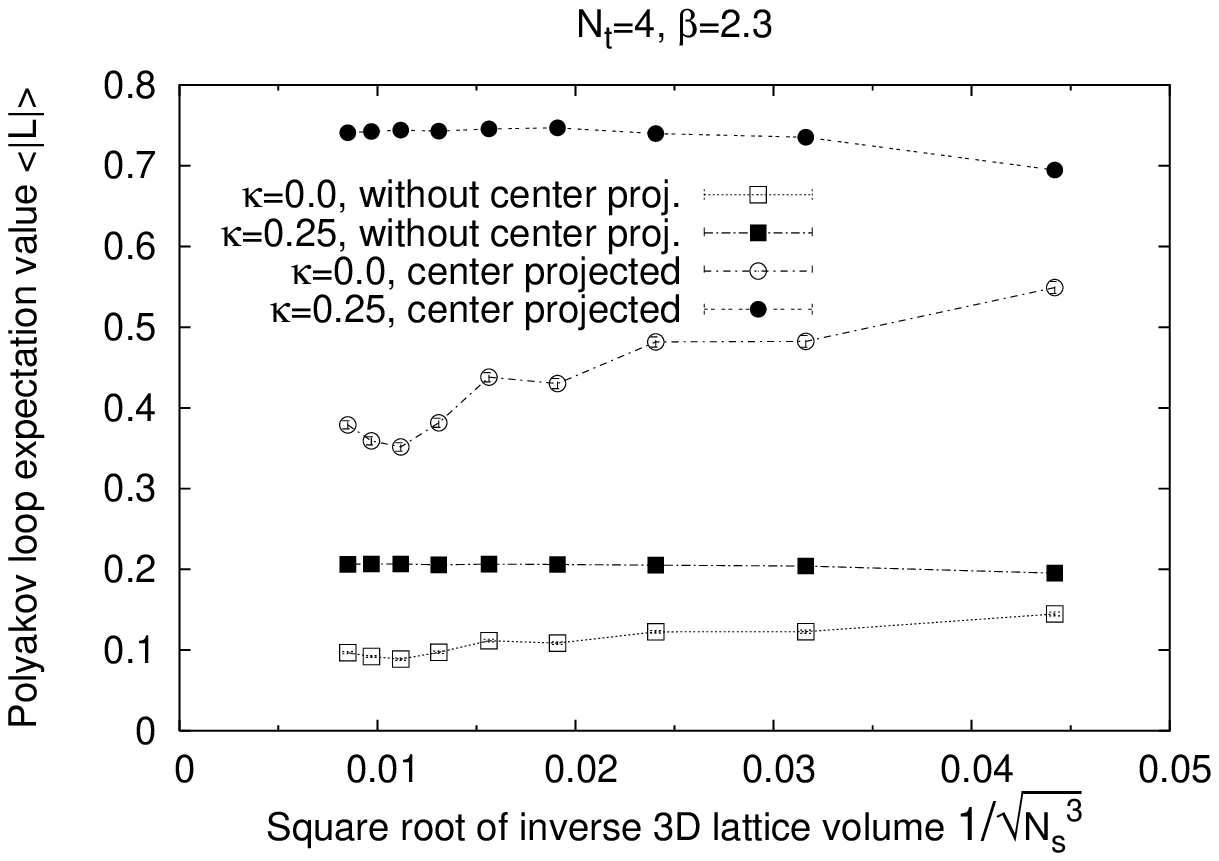}
  \includegraphics[width=1.0\linewidth,height=0.4\textheight,keepaspectratio]{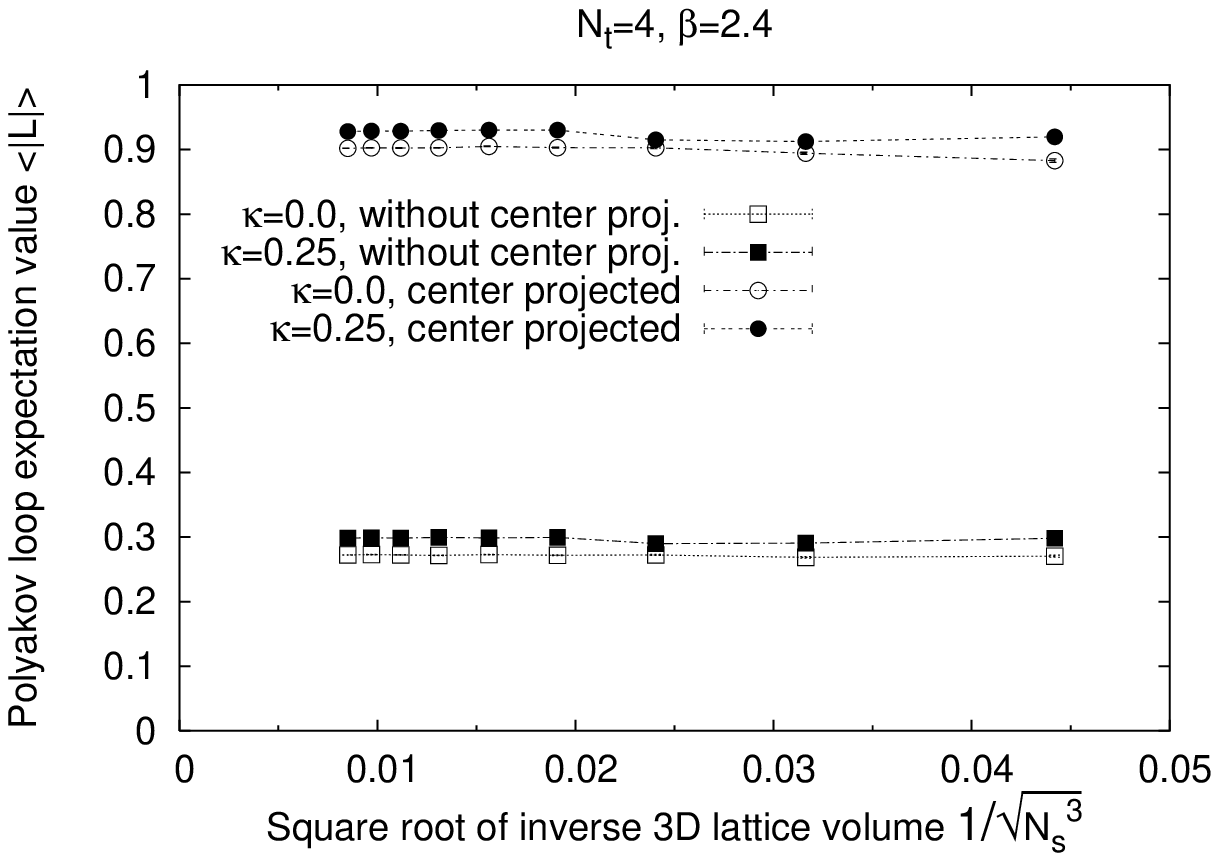}
  \caption{Expectation values of the Polyakov loop in the deconfined phase.
    Open symbols denote measurements at $\kappa=0.0$, filled symbols at $\kappa=0.25$.
    Squares are used for measurements on unprojected, full SU(2) configurations and circles for center projected fields.}
  \label{fig:polyakovdeconf}
\end{figure}
\begin{figure}
  \includegraphics[width=1.0\linewidth,height=0.4\textheight,keepaspectratio]{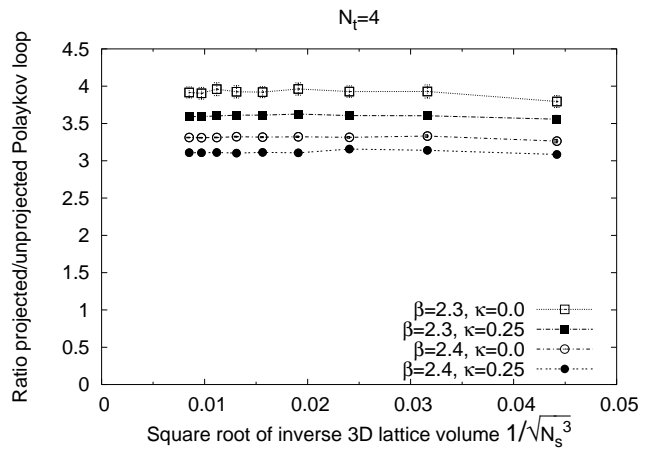}
  \caption{Ratio of projected to unprojected Polyakov loop expectation values in the deconfined phase.
    Open symbols denote measurements at $\kappa=0.0$, filled symbols at $\kappa=0.25$.}
  \label{fig:polyakovratiodeconf}
\end{figure}
The inclusion of a massive scalar field does not lead to a qualitative change of the system in the high temperature phase.
This is reflected by our results for $\pol$ and $\zpol$ at $\beta=2.3$ and $2.4$, seen in \fig{fig:polyakovdeconf} and \fig{fig:polyakovratiodeconf}.
Again the projected and the unprojected measurements are consistent.
As in the ``confined'' phase, the projected values are higher than the unprojected ones, but their ratio is almost constant with respect to $N_s$.
There is some $\kappa$ dependence in projected and unprojected Polyakov lines right above the high $T$ phase transition (which is just below $\beta=2.3$ at $N_t=4$), but the ratios of projected and unprojected lines are nevertheless almost constant, and the $\kappa$ dependence in $\pol$ and $\zpol$ is rather small at still higher temperatures in the deconfined regime.

From these calculations we can conclude that we see center dominance for Polyakov loops also in the presence of scalar matter fields, both above and below the finite temperature phase transition.
This means that $\pol$ and $\zpol$ are in the infinite volume limit either both zero, or both non-zero, at any temperature with or without dynamical matter fields.
Once again, the long range physics of the system seems to be encoded in the center degrees of freedom.
In particular, the presence of a dynamical matter field must alter the spatial distribution of P-vortices, in such a way that P-vortex fluctuations no longer bring $\zpol$ to zero in the large volume limit, even in the low temperature regime.

\section{\label{sec:zeroT}Zero Temperature}

Our results for $T=0$ have been obtained on hypercubic $10^4$ lattices for $\beta \le 1.9$, on $12^4$ lattices for $\beta = 2.1$, and on $16^4$ lattices at $\beta = 2.3$.
For $\beta = 0.25$, we have performed a detailed scan over $\kappa$ on a $16^4$ lattice.
Each measurement has been done at $\lambda=1.0$ using 800 configurations.
Otherwise, we have used the same parameters and methods as for the simulations at finite temperature.

Due to the small overlap of flux-tube states with gluelump states, it is difficult to measure string breaking in the ``confined'' phase using only Wilson loops, since the breaking (and hence perimeter-law behavior) occurs only for rather large loops, where extremely high statistics are needed.%
\footnote{Nevertheless, we note that recently adjoint string breaking has been successfully measured using Wilson loops in ref.\ \cite{kf03a}, using a powerful algorithm for noise reduction devised by L\"uscher and Weisz \cite{lw01a}.}
This problem is reflected in the vortex model.
Two of us (R.B.\ and M.F.) have carefully investigated the Wilson loop observable in a wide range of the phase diagram ($1.0 \le \beta \le 2.3, 0.0 \le \kappa \le 1.1$) on lattices up to $22^4$ \cite{bf02a}.
As in the unprojected, full system, the influence of the Higgs on the Wilson loop observable $W(R,T)$ can be effectively described --- for our available small loops $R,T \le 4$ --- by some shift $\beta \rightarrow \beta_{eff}$.
Center dominance holds again.

\begin{figure}
  \includegraphics[width=1.0\columnwidth]{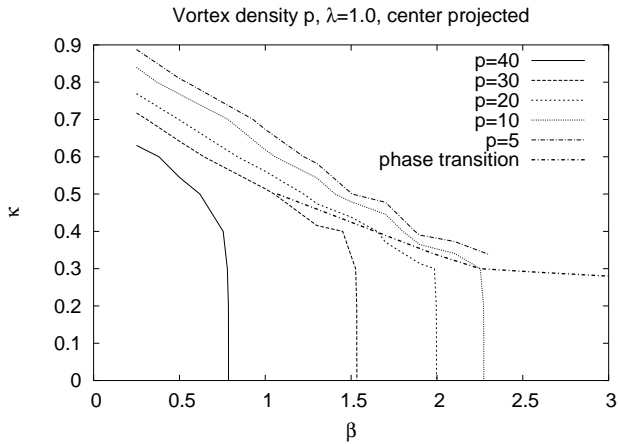}
  \caption{P-vortex density $p$ in the $\beta-\kappa$ plane.
    Lattice sizes are indicated in the text.  The phase transition
    line is taken from \cite{jlnv85a}.}
  \label{fig:vortexfraccontour}
\end{figure}

At the transition to the Higgs phase, the string tension measured by Wilson loops disappears even for small distances.
We have found that the phase transition line in the gauge-Higgs model is closely correlated, even for small lattices, with a rapid drop in the vortex density $p$, as shown in \fig{fig:vortexfraccontour}.
In this figure we plot the density of P-vortex plaquettes in percent as contours in the $\beta-\kappa$ plane.
Data for the phase transition line are taken from \cite{jlnv85a}.
In the ``confined'' phase, the vortex density is almost independent of $\kappa$, whereas the vortex density decreases rapidly, with increasing $\kappa$, in the neighborhood of the phase transition line.
This rather sudden decrease in vortex density with $\kappa$, at fixed $\beta$ was first observed by Langfeld \cite{la02a}, and by two of us (R.B.\ and M.F.) \cite{bf02a}.
It is interesting that this region of rapid decrease in density, seen in \fig{fig:vortexfraccontour}, appears to extend beyond the actual thermodynamic transition line, all the way to the $\beta=0$ axis.
We may ask whether this is simply crossover behavior, or an indication of some more interesting phenomenon.

\subsection{The Kert\'esz Line}

A ``Kert\'esz line'', originally discovered in Ising spin systems, is a line of percolation transitions which is free of any non-analyticity in the corresponding free energy.
This is therefore \emph{not} a line of thermodynamic phase transitions, as usually defined.
It was originally suggested by Langfeld \cite{la02a} that the rapid decrease of P-vortex density in the gauge-Higgs theory, in a region where there is no thermodynamic transition, might also be associated with a Kert\'esz line of P-vortex percolation transitions.
We will now report on some results which support that idea.

Our simulations are performed on $16^4$ lattices at $\beta=0.25$, which is far below the end of the thermodynamic phase transition line around $\beta=1$.
In \fig{fig:B25} we display the following observables:
\begin{figure}
  \includegraphics[width=1.0\columnwidth]{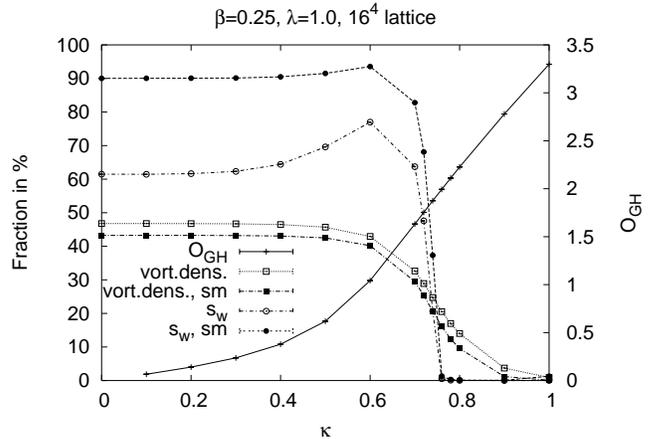}
  \caption{P-vortex density $p$, weighted average vortex size $s_w$, and the one-link gauge-Higgs energy density $\mathrm{O}_\mathrm{GH}$ for $\beta=0.25$.
    Open symbols denote measurements after center projection, for filled symbols (labelled with ``sm'') elementary vortex cubes are removed in addition.}
  \label{fig:B25}
\end{figure}

\begin{itemize}
\item The one-link contribution to the gauge-Higgs energy density
  \begin{equation}
    \mathrm{O}_\mathrm{GH}:=\big\langle\mathrm{Re}\left[\Phi^\dagger(x)
    U_\mu(x) \Phi(x+\hat{\mu})\right]\big\rangle.
  \end{equation}
  This is a local, thermodynamic observable.
\item The P-vortex density $p$.
\item The weighted average vortex size $s_w$. This is the relative size $||c_{i(a)}||/N_p$ of the vortex $c_{i(a)}$ containing the P-plaquette $P_a$, averaged over all $N_P$ P-plaquettes in the projected gauge field configuration%
\footnote{This observable resembles the average cluster size $S$ of percolation theory.
In our case the clusters are the vortices.
$S$ differs from our $s_w$ in that i) it is not the average absolute size, but the average fraction with respect to all P-plaquettes;
and ii) all clusters resp.\ vortices, including the percolating one, are included in the sum.}:
  \begin{equation}
    s_w := 
    \frac{1}{N_P} \sum_{a=1}^{N_P} \frac{||c_{i(a)}||}{N_P} =
    \sum_{i=1}^{N_v} \frac{||c_i||^2}{N_P^2}\;.
    \label{eq:avvortsiz}
  \end{equation}
  Here $||c_i||$ is the number of plaquettes in vortex $c_i$, $i(a)$ the index of the vortex containing plaquette $P_a$, and $N_v$ is the number of vortices in the projected configuration.
In the limit of infinite lattice volume and finite vortex density, $s_w=1$ if all P-plaquettes belong to a single percolating vortex, and $s_w=0$ for the depercolation case.
  This is a highly non-local observable, and it is the quantity which we use to detect a percolation transition.
\end{itemize}

At $\beta=0.25$, it can be seen from \fig{fig:B25} that the one-link energy density $\mathrm{O}_\mathrm{GH}$ is a smooth function of $\kappa$, and displays no evidence of a transition.
The vortex density, although dropping rapidly in the interval $\kappa \in [0.6,0.9]$, also shows no sign of any discontinuity.

The vortex observables in \fig{fig:B25} have been computed in two ways.
Open symbols denote measurements on center projected configurations.
For filled symbols (labelled with ``sm''), the first step of a vortex smoothing procedure, described in ref.\ \cite{bfgo99a}, has been applied.
This first step completely removes the smallest (one-cube) vortices consisting of six P-plaquettes.
We have checked that down to $\beta=1.0$ this removal does not alter projected Creutz ratios $\chi(R,R)$, $R \geq 2$.
The smoothing procedure decreases the vortex density $p$ depicted in \fig{fig:B25} slightly, but does not qualitatively change the shape of the curve.
Smoothed or not, the vortex density decreases continuously from just under 50\% to zero as $\kappa$ increases, with no evidence of any sudden change in phase.

There is, however, a clear sign of a sudden transition, around $\kappa=0.74$, in the weighted average vortex size $s_w$.
The plot of $s_w$ in \fig{fig:vortexfraccontour} shows a sharp depercolation transition for this observable.
For $\kappa \leq 0.7$, the majority of P-plaquettes belong to one big, percolating vortex surface, but for $\kappa \geq 0.74$, this surface is split into many small vortices.
This is even more pronounced if we remove  the smallest (one cube) vortices using the smoothing procedure.
After smoothing, but below the transition, the average vortex contains over 90\% of all P-plaquettes.
Just after the transition, this number drops to under 1.5\%.
This is a fairly convincing sign of the existence of a percolation transition.
We have also checked for percolation directly, by measuring the spatial extension of the largest vortex surface on the lattice.
In the Higgs phase, the largest vortex always fits inside a hypercube which is smaller than the full lattice, while in the ``confined'' phase this is not possible;
the largest connected vortex surface extends through the entire lattice, irrespective of lattice size.
Thus we find center vortex percolation in the ``confined'' phase of the gauge-Higgs theory, and no percolation in the Higgs phase.
There is a first-order phase transition line, which has an endpoint in the interior of the $\beta-\kappa$ plane but then continues as a Kert\'esz line, completely separating these two regions of the phase diagram.

It is quite natural that there should be no vortex percolation in the Higgs phase, since percolation is a necessary condition for achieving an area law for Wilson loops, which of course is absent in this phase.
But an asymptotic string tension is \emph{also} absent in the ``confined'' phase, due to color screening, yet in this case we do have vortex percolation.
This is not a paradox, however.
The vortex confinement mechanism  requires that vortex piercings of the minimal area of a large Wilson loop are uncorrelated, and percolation alone is not enough to ensure this property.
As an example, consider a percolating P-vortex surface having the form of a branched polymer (a form which often arises in numerical simulations of random surfaces).
In this case each piercing of a plane surface will be accompanied by a second nearby piercing, with the pair contributing no net center flux, and therefore no net disordering, to the Wilson loop.
In this example there is no confinement, even though the P-vortex percolates throughout the lattice.
Percolation is therefore a necessary, but not a sufficient condition for confinement \cite{gr02a}.

\section{\label{sec:concl}Conclusions}

We have observed center dominance in SU(2) gauge-Higgs theory for the Polyakov line observable, both at low and high temperatures, throughout the $\beta-\kappa$ coupling plane.
Polyakov line expectation values $\pol$ and $\zpol$, on the full and projected lattices respectively, are either both zero, or both non-zero, in the infinite volume limit.

 There is no true confining phase in a gauge-Higgs theory at $\kappa > 0$, or in any gauge theory with matter in the fundamental representation of the gauge group.
Nevertheless, we have found a non-local observable, sensitive to vortex percolation, which distinguishes between the confinement-like and Higgs regions of the phase diagram.
The confinement-like region is characterized by an area-law decay of Wilson loops up to a certain string-breaking scale, while in the Higgs-like region there is no area-law falloff at any scale.
The two regions are separated by a line of thermodynamic first order transitions, which turns into a Kert\'esz line at $\beta<1$.
As in the Ising spin system, the Kert\'esz line is a line of percolation transitions, with the free energy analytic across the transition.
In the confinement-like region, center vortices percolate throughout the lattice.
In the Higgs region, they do not.

The sharp transition between the Higgs and confinement-like regions has been seen in other ways.
Langfeld \cite{la02a} has noted that after fixing to Landau gauge, there is remnant unfixed global symmetry, and that this symmetry is unbroken in the confinement-like phase, and broken across the Kert\'esz line in the Higgs phase.
A similar observation, this time in Coulomb gauge, has been made very recently by two of us (J.G.\ and {\v S}.O.) and Zwanziger in ref.\ \cite{go03a,goz03a}.
It should be noted that the order parameter for remnant symmetry breaking, if expressed as a gauge-invariant observable, is highly non-local, as is the order parameter for percolation.

The fact that vortices percolate in the confinement-like region of the gauge-Higgs phase diagram, yet the asymptotic string tension is zero, demonstrates that while vortex percolation is a necessary condition for confinement, it is not also a sufficient condition.
Vortex percolation without confinement implies that vortex piercings of a plane are not entirely random, but are paired in some way.
An example of such a percolating but non-confining surface is a branched polymer, but the actual structure of percolating center vortices, in the confinement-like region of gauge-Higgs theory, is not yet known.

\begin{acknowledgments}
Our research is supported in part by Fonds zur F\"orderung der Wissenschaftlichen Forschung P13997-TPH and P14435-TPH (R.B.\ and M.F.), the U.S.\ Department of Energy under Grant No.\ DE-FG03-92ER40711 (J.G.), and the Slovak Grant Agency for Science, Grant No.\ 2/3106/2003 (\v{S}.O.).
\end{acknowledgments}

\bibliographystyle{apsrev}
\bibliography{bertle}

\begin{thebibliography}{21}
\expandafter\ifx\csname natexlab\endcsname\relax\def\natexlab#1{#1}\fi
\expandafter\ifx\csname bibnamefont\endcsname\relax
  \def\bibnamefont#1{#1}\fi
\expandafter\ifx\csname bibfnamefont\endcsname\relax
  \def\bibfnamefont#1{#1}\fi
\expandafter\ifx\csname citenamefont\endcsname\relax
  \def\citenamefont#1{#1}\fi
\expandafter\ifx\csname url\endcsname\relax
  \def\url#1{\texttt{#1}}\fi
\expandafter\ifx\csname urlprefix\endcsname\relax\def\urlprefix{URL }\fi
\providecommand{\bibinfo}[2]{#2}
\providecommand{\eprint}[2][]{\url{#2}}

\bibitem[{\citenamefont{Greensite}(2003)}]{gr03a}
\bibinfo{author}{\bibfnamefont{J.}~\bibnamefont{Greensite}},
  \bibinfo{journal}{Prog. Part. Nucl. Phys.} \textbf{\bibinfo{volume}{51}},
  \bibinfo{pages}{1} (\bibinfo{year}{2003}), \eprint{hep-lat/0301023}.

\bibitem[{\citenamefont{Knechtli and Sommer}(1998)}]{ks98a}
\bibinfo{author}{\bibfnamefont{F.}~\bibnamefont{Knechtli}} \bibnamefont{and}
  \bibinfo{author}{\bibfnamefont{R.}~\bibnamefont{Sommer}}
  (\bibinfo{collaboration}{ALPHA collaboration}), \bibinfo{journal}{Phys.
  Lett.} \textbf{\bibinfo{volume}{B440}}, \bibinfo{pages}{345}
  (\bibinfo{year}{1998}), \eprint{hep-lat/9807022}.

\bibitem[{\citenamefont{Fradkin and Shenker}(1979)}]{fs79a}
\bibinfo{author}{\bibfnamefont{E.}~\bibnamefont{Fradkin}} \bibnamefont{and}
  \bibinfo{author}{\bibfnamefont{S.~H.} \bibnamefont{Shenker}},
  \bibinfo{journal}{Phys. Rev.} \textbf{\bibinfo{volume}{D19}},
  \bibinfo{pages}{3682} (\bibinfo{year}{1979}).

\bibitem[{\citenamefont{Mack and Meyer-Ortmanns}(1982)}]{mm82a}
\bibinfo{author}{\bibfnamefont{G.}~\bibnamefont{Mack}} \bibnamefont{and}
  \bibinfo{author}{\bibfnamefont{H.}~\bibnamefont{Meyer-Ortmanns}},
  \bibinfo{journal}{Nucl. Phys.} \textbf{\bibinfo{volume}{B200}},
  \bibinfo{pages}{249} (\bibinfo{year}{1982}).

\bibitem[{\citenamefont{Meyer-Ortmanns}(1984)}]{me84b}
\bibinfo{author}{\bibfnamefont{H.}~\bibnamefont{Meyer-Ortmanns}},
  \bibinfo{journal}{Nucl. Phys.} \textbf{\bibinfo{volume}{B235}},
  \bibinfo{pages}{115} (\bibinfo{year}{1984}).

\bibitem[{\citenamefont{{Del Debbio} et~al.}(1998)\citenamefont{{Del Debbio},
  Faber, Giedt, Greensite, and {Olejn{\'{\i}}k}}}]{dfggo98a}
\bibinfo{author}{\bibfnamefont{L.}~\bibnamefont{{Del Debbio}}},
  \bibinfo{author}{\bibfnamefont{M.}~\bibnamefont{Faber}},
  \bibinfo{author}{\bibfnamefont{J.}~\bibnamefont{Giedt}},
  \bibinfo{author}{\bibfnamefont{J.}~\bibnamefont{Greensite}},
  \bibnamefont{and} \bibinfo{author}{\bibfnamefont{{\v
  S}.}~\bibnamefont{{Olejn{\'{\i}}k}}}, \bibinfo{journal}{Phys. Rev.}
  \textbf{\bibinfo{volume}{D58}}, \bibinfo{pages}{094501}
  (\bibinfo{year}{1998}), \eprint{hep-lat/9801027}.

\bibitem[{\citenamefont{{Del Debbio} et~al.}(1997)\citenamefont{{Del Debbio},
  Faber, Greensite, and {Olejn{\'{\i}}k}}}]{dfgo97b}
\bibinfo{author}{\bibfnamefont{L.}~\bibnamefont{{Del Debbio}}},
  \bibinfo{author}{\bibfnamefont{M.}~\bibnamefont{Faber}},
  \bibinfo{author}{\bibfnamefont{J.}~\bibnamefont{Greensite}},
  \bibnamefont{and} \bibinfo{author}{\bibfnamefont{{\v
  S}.}~\bibnamefont{{Olejn{\'{\i}}k}}}, \bibinfo{journal}{Phys. Rev.}
  \textbf{\bibinfo{volume}{D55}}, \bibinfo{pages}{2298} (\bibinfo{year}{1997}),
  \eprint{hep-lat/9610005}.

\bibitem[{\citenamefont{Chernodub et~al.}(1999)\citenamefont{Chernodub,
  Polikarpov, Veselov, and Zubkov}}]{cpvz98a}
\bibinfo{author}{\bibfnamefont{M.~N.} \bibnamefont{Chernodub}},
  \bibinfo{author}{\bibfnamefont{M.~I.} \bibnamefont{Polikarpov}},
  \bibinfo{author}{\bibfnamefont{A.~I.} \bibnamefont{Veselov}},
  \bibnamefont{and} \bibinfo{author}{\bibfnamefont{M.~A.}
  \bibnamefont{Zubkov}}, \bibinfo{journal}{Nucl. Phys. Proc. Suppl.}
  \textbf{\bibinfo{volume}{73}}, \bibinfo{pages}{575} (\bibinfo{year}{1999}),
  \eprint[http://arXiv.org/abs]{hep-lat/9809158}.

\bibitem[{\citenamefont{Langfeld et~al.}(1999)\citenamefont{Langfeld, Tennert,
  Engelhardt, and Reinhardt}}]{lter99a}
\bibinfo{author}{\bibfnamefont{K.}~\bibnamefont{Langfeld}},
  \bibinfo{author}{\bibfnamefont{O.}~\bibnamefont{Tennert}},
  \bibinfo{author}{\bibfnamefont{M.}~\bibnamefont{Engelhardt}},
  \bibnamefont{and}
  \bibinfo{author}{\bibfnamefont{H.}~\bibnamefont{Reinhardt}},
  \bibinfo{journal}{Phys. Lett.} \textbf{\bibinfo{volume}{B452}},
  \bibinfo{pages}{301} (\bibinfo{year}{1999}), \eprint{hep-lat/9805002}.

\bibitem[{\citenamefont{Engelhardt et~al.}(2000)\citenamefont{Engelhardt,
  Langfeld, Reinhardt, and Tennert}}]{elrt99a}
\bibinfo{author}{\bibfnamefont{M.}~\bibnamefont{Engelhardt}},
  \bibinfo{author}{\bibfnamefont{K.}~\bibnamefont{Langfeld}},
  \bibinfo{author}{\bibfnamefont{H.}~\bibnamefont{Reinhardt}},
  \bibnamefont{and} \bibinfo{author}{\bibfnamefont{O.}~\bibnamefont{Tennert}},
  \bibinfo{journal}{Phys. Rev.} \textbf{\bibinfo{volume}{D61}},
  \bibinfo{pages}{054504} (\bibinfo{year}{2000}), \eprint{hep-lat/9904004}.

\bibitem[{\citenamefont{Bertle et~al.}(1999)\citenamefont{Bertle, Faber,
  Greensite, and {Olejn{\'{\i}}k}}}]{bfgo99a}
\bibinfo{author}{\bibfnamefont{R.}~\bibnamefont{Bertle}},
  \bibinfo{author}{\bibfnamefont{M.}~\bibnamefont{Faber}},
  \bibinfo{author}{\bibfnamefont{J.}~\bibnamefont{Greensite}},
  \bibnamefont{and} \bibinfo{author}{\bibfnamefont{{\v
  S}.}~\bibnamefont{{Olejn{\'{\i}}k}}}, \bibinfo{journal}{J. High Energy Phys.}
  \textbf{\bibinfo{volume}{03}}, \bibinfo{pages}{019} (\bibinfo{year}{1999}),
  \eprint{hep-lat/9903023}.

\bibitem[{\citenamefont{Langfeld}(2002)}]{la02a}
\bibinfo{author}{\bibfnamefont{K.}~\bibnamefont{Langfeld}},
  \bibinfo{journal}{to appear in proc. Strong and Electroweak Matter}
  (\bibinfo{year}{2002}), \eprint{hep-lat/0212032}.

\bibitem[{\citenamefont{Bertle and Faber}(2002)}]{bf02a}
\bibinfo{author}{\bibfnamefont{R.}~\bibnamefont{Bertle}} \bibnamefont{and}
  \bibinfo{author}{\bibfnamefont{M.}~\bibnamefont{Faber}}, \bibinfo{journal}{to
  appear in proc. Confinement V}  (\bibinfo{year}{2002}),
  \eprint{hep-lat/0212027}.

\bibitem[{\citenamefont{Kert{\'e}sz}(1989)}]{ke89a}
\bibinfo{author}{\bibfnamefont{J.}~\bibnamefont{Kert{\'e}sz}},
  \bibinfo{journal}{Physica} \textbf{\bibinfo{volume}{A161}},
  \bibinfo{pages}{58} (\bibinfo{year}{1989}).

\bibitem[{\citenamefont{Satz}(2002)}]{sa02a}
\bibinfo{author}{\bibfnamefont{H.}~\bibnamefont{Satz}},
  \bibinfo{journal}{Comput. Phys. Commun.} \textbf{\bibinfo{volume}{147}},
  \bibinfo{pages}{46} (\bibinfo{year}{2002}), \eprint{hep-lat/0110013}.

\bibitem[{\citenamefont{Kratochvila and de~Forcrand}(2003)}]{kf03a}
\bibinfo{author}{\bibfnamefont{S.}~\bibnamefont{Kratochvila}} \bibnamefont{and}
  \bibinfo{author}{\bibfnamefont{P.}~\bibnamefont{de~Forcrand}},
  \bibinfo{journal}{CERN-TH-2003-119}  (\bibinfo{year}{2003}),
  \eprint{hep-lat/0306011}.

\bibitem[{\citenamefont{L{\"u}scher and Weisz}(2001)}]{lw01a}
\bibinfo{author}{\bibfnamefont{M.}~\bibnamefont{L{\"u}scher}} \bibnamefont{and}
  \bibinfo{author}{\bibfnamefont{P.}~\bibnamefont{Weisz}},
  \bibinfo{journal}{JHEP} \textbf{\bibinfo{volume}{09}}, \bibinfo{pages}{010}
  (\bibinfo{year}{2001}), \eprint{hep-lat/0108014}.

\bibitem[{\citenamefont{Jers{\'a}k et~al.}(1985)\citenamefont{Jers{\'a}k, Lang,
  Neuhaus, and Vones}}]{jlnv85a}
\bibinfo{author}{\bibfnamefont{J.}~\bibnamefont{Jers{\'a}k}},
  \bibinfo{author}{\bibfnamefont{C.~B.} \bibnamefont{Lang}},
  \bibinfo{author}{\bibfnamefont{T.}~\bibnamefont{Neuhaus}}, \bibnamefont{and}
  \bibinfo{author}{\bibfnamefont{G.}~\bibnamefont{Vones}},
  \bibinfo{journal}{Phys. Rev.} \textbf{\bibinfo{volume}{D32}},
  \bibinfo{pages}{2761} (\bibinfo{year}{1985}).

\bibitem[{\citenamefont{Gliozzi and Rago}(2002)}]{gr02a}
\bibinfo{author}{\bibfnamefont{F.}~\bibnamefont{Gliozzi}} \bibnamefont{and}
  \bibinfo{author}{\bibfnamefont{A.}~\bibnamefont{Rago}},
  \bibinfo{journal}{Phys. Rev.} \textbf{\bibinfo{volume}{D66}},
  \bibinfo{pages}{074511} (\bibinfo{year}{2002}), \eprint{hep-lat/0206017}.

\bibitem[{\citenamefont{Greensite et~al.}(2003)\citenamefont{Greensite,
  {Olejn{\'{\i}}k}, and Zwanziger}}]{goz03a}
\bibinfo{author}{\bibfnamefont{J.}~\bibnamefont{Greensite}},
  \bibinfo{author}{\bibfnamefont{{\v S}.}~\bibnamefont{{Olejn{\'{\i}}k}}},
  \bibnamefont{and}
  \bibinfo{author}{\bibfnamefont{D.}~\bibnamefont{Zwanziger}},
  \bibinfo{journal}{in preparation}  (\bibinfo{year}{2003}).

\bibitem[{\citenamefont{Greensite and {Olejn{\'{\i}}k}}(2003)}]{go03a}
\bibinfo{author}{\bibfnamefont{J.}~\bibnamefont{Greensite}} \bibnamefont{and}
  \bibinfo{author}{\bibfnamefont{{\v S}.}~\bibnamefont{{Olejn{\'{\i}}k}}}
  (\bibinfo{year}{2003}), \eprint{hep-lat/0309172}.

\end{thebibliography}

\end{document}